# Direct detonation initiation and propagation in methane/air mixtures containing coal particles

*Shengnan Li[1,2], Shangpeng Li[2], Shumeng Xie[2], Yong Xu[2], Ke Gao[1], Huangwei Zhang[2,*]*

[1] *College of Safety Science and Engineering, Liaoning Technical University, Huludao 125105, China*
[2] *Department of Mechanical Engineering, National University of Singapore, 9 Engineering Drive 1, Singapore, 117576, Republic of Singapore*

**Abstract**

The mechanisms of direct detonation initiation (DDI) in methane/air mixtures containing coal particles are investigated through simulations conducted using the Eulerian-Lagrangian method in a two-dimensional configuration. Methane-air combustion is modelled with a detailed chemical mechanism involving 36 species and 219 reactions, while coal particle surface reactions are computed using a kinetic/diffusion-limited rate model. The findings indicate that shock waves generated from the hotspot can initiate detonation through heterogeneous and homogeneous reactions, with contributions from both methane and particle combustion. Coal particle surface reactions provide the dominant energy for detonation initiation, whereas gas-phase reactions enhance detonation stability during propagation. The difficulty of achieving detonation initiation exhibits a non-linear dependence on particle concentrations and gas equivalence ratios. An optimal particle concentration and gas equivalence ratio for successful DDI is identified. Smaller particles are found to facilitate detonation initiation more effectively. Key processes in DDI of two-phase mixtures are identified, including particle heating, methane combustion, and particle burning. Three DDI modes—critical, stable, and cell-free—are observed based on particle concentration. As particle concentration increases, the temperatures of both particles and gas become close, initially rising and then decreasing with further increases in particle concentration. Additionally, the introduction of coal particles gives rise to two distinct stages in gas-phase reactions.

**Novelty and significance statement**

This numerical study offers new insights into the properties of direct detonation initiation in methane/coal particle/air mixtures. Criteria for detonation identification are established to evaluate the direct detonation initiation phenomenon under the influence of multiple factors. Furthermore, the study highlights the varying roles of methane and coal particle reactions in hybrid detonation initiation and propagation. Several characteristic structures in coal particle/air mixtures are identified for the first time. These findings contribute valuable knowledge to coal mine safety prevention and control efforts.

**CRediT authorship contribution statement**

**Shengnan Li**: Conceptualization, Methodology, Writing – original draft. **Shangpeng Li**: Writing – review & editing, Methodology, Formal analysis. **Shumeng Xie**: Methodology, investigation, Software. **Yong Xu**: Methodology, Software. **Ke Gao**: Supervision, Funding acquisition. **Huangwei Zhang**: Conceptualization, Resources, Writing – review & editing.

# Nomenclature

## *Terminology*

| | | | | |
|---|---|---|---|---|
| $t$ | Time [s] | | $h_c$ | Convective heat transfer coefficient [W/m²/K] |
| $p$ | Pressure [Pa] | | $\dot{Q}_{hm}$ | Heat release rate of homogeneous reactions [J/m³/s] |
| $p_0$ | Initial pressure [Pa] | | $\dot{Q}_{ht}$ | Heat release rate of heterogeneous reactions [J/m³/s] |
| $T$ | Gas temperature [K] | | $A_p$ | Particle surface area [m²] |
| $T_0$ | Initial gas temperature [K] | | $\mathbf{u}$ | Gas velocity [m/s] |
| $R$ | Universal gas constant [J/mol/K] | | $\mathbf{u}_p$ | Particle velocity [m/s] |
| $\nabla p$ | Pressure gradient [Pa/m] | | $R_k$ | Kinetic rate coefficient |
| $E_a$ | Activation energy [kJ/mol] | | $\mathbf{F}_d$ | Particle drag force [N] |
| $m_p$ | Particle mass [kg] | | $\mathbf{F}_p$ | Pressure gradient force [N] |
| $\dot{m}_p$ | Particle surface reaction rate [kg/s] | | $Re_p$ | Particle Reynolds number |
| $T_p$ | Particle temperature [K] | | $T_s$ | Hotspot temperature [K] |
| $d_p$ | Particle diameter [m] | | $p_s$ | Hotspot pressure [Pa] |
| $D_o$ | Diffusion rate coefficient | | $D_{CJ}$ | C—J detonation speed [m/s] |
| $p_{ox}$ | Partial pressure of oxidant species [Pa] | | $D_{LSF}$ | Leading shock front speed [m/s] |
| $\phi$ | Equivalence ratio of gas mixtures | | $V_p$ | Volume of a single particle [m³] |
| $c_0$ | Particle concentration [kg/m³] | | $\tau_1$ | Overdriven detonation [s] |
| $R_c$ | Cellular detonation radius | | $\tau_2$ | developing cellular detonation [s] |
| $\dot{Q}_c$ | Convective heat transfer rate [J/s] | | $c_{p,p}$ | Particle heat capacity at constant pressure [J/kg/K] |

## *Greek letter*

| | | | | |
|---|---|---|---|---|
| $\rho$ | Gas density [kg/m³] | | $\mu$ | Dynamic viscosity [kg/m/s] |

## *Acronym*

| | | | | |
|---|---|---|---|---|
| LSF | Leading shock front | | RF | Reaction front |
| DW | Detonation wave | | HRL | Half-reaction length [m] |
| C-J | Chapman–Jouguet | | ZND | Zel'dovich-von Neumann-Döring |
| IW | Incident wave | | MS | Mach stem |
| TP | Triple point | | DDI | Direct detonation initiation |
| TW | Transverse wave | | RT | Rayleigh–Taylor |

# 1. Introduction

In recent years, explosion accidents involving methane/coal particle/air mixtures have increased significantly, highlighting their critical safety implications. Underground coal mines, in particular, are highly vulnerable to explosions caused by hybrid mixtures [1]. Meanwhile, detonation has been tested in propulsion systems to harness the resulting pressure gain, with recent experiments in gas-solid rotating detonation engines using pulverised solid fuels [2]. However, understanding of combustion and explosion dynamics in methane and coal particle mixtures remains limited.

Direct detonation initiation (DDI) refers to the instantaneous formation of detonation without the pre-detonation stage of flame acceleration [3]. Shen et al. [4] found that the critical energy required for successful DDI depends on the stability of detonation waves. As the critical energies decrease, detonation transitions through three phases: stable, mildly unstable, and highly unstable. Zhang et al. [5] highlighted the significant impact of the gas equivalence ratio on DDI in typical gaseous fuels (e.g., $C_2H_2$, $C_2H_4$, $C_3H_8$, $H_2$) and oxygen mixtures. They observed an increase in critical energy up to a threshold equivalence ratio, beyond which it declined. Notably, the dynamics of DDI in an open environment differ significantly from those during the transition from tubes to open spaces [6–8]. In a typical DDI process, the ignition source generates sufficiently strong blast waves that can directly initiate the detonation wave without a flame acceleration stage. This type of detonation wave, whether spherical or cylindrical, is not constrained by wall conditions during propagation. The surface area continues to increase, forming a divergent detonation wave, making the propagation mechanism of explosion waves in straight tubes inapplicable.

Previous studies on DDI have primarily focused on gaseous mixtures, where detonation is mainly driven by rapid chemical reactions and shock wave propagation through a uniformly heated and energised gas medium. However, the introduction of solid particles leads to significant differences due to the interactions between the gas and particles, as illustrated in Fig. 1. For example, the presence of particles can affect the structure of shock and detonation waves. The transfer of mass, energy, and momentum between coal particles and surrounding gas alters detonation velocity and stability. These interactions become more complex at higher particle concentrations. Factors such as particle heating, surface reactions, distribution characteristics, and gas-particle interactions substantially influence DDI in two-phase mixtures. These result in distinct characteristics, such as shock/detonation speed, detonation cell size, and overpressure, when compared to gaseous detonations (see Fig. 1). In hybrid detonations involving both gaseous and solid fuels, the leading shock and reaction fronts are closely coupled, with the shock front compressing the mixture and sustaining ignition at the reaction front [9]. The motion, heating, and surface reactions of the coal particles contribute to mass, momentum, and energy transfers, including the thermo-mechanical-chemical interactions between coal particles and the flammable gas [10].

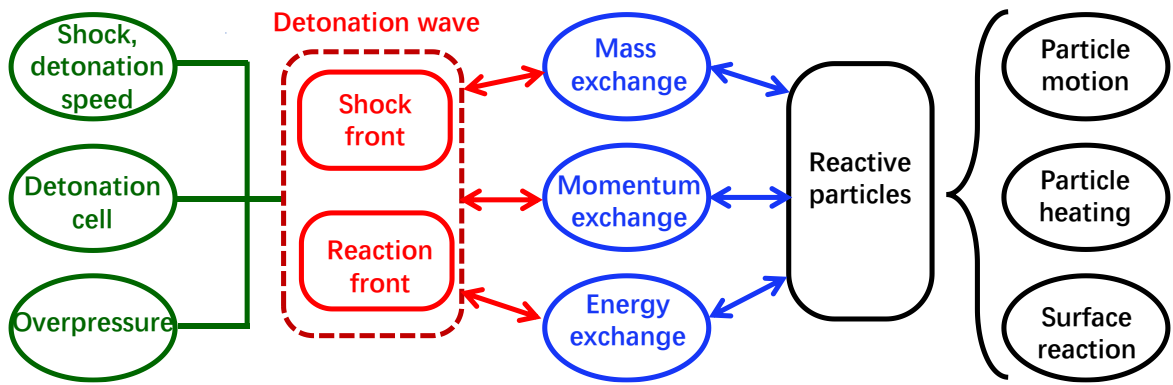

Figure 1: Schematic of DDI of coal particle/methane/ air mixtures.

To the best of our knowledge, research on DDI in gas/solid two-phase mixtures, particularly those involving methane and coal particles as the detonated medium, remains limited. This study presents detailed simulations to investigate the direct initiation and propagation of two-phase detonation. Firstly, we examine the effects of particle concentration, particle diameter, and gas equivalence ratio on DDI. Secondly, we analyse the detonation structure and DDI mode under various conditions. Finally, we explore the interactions and heat (mass) transfer mechanisms in two-phase detonations. This paper is structured as follows: Section 2 introduces the physical and mathematical models, along with the simulation cases. Results and discussion are presented in Sections 3 and 4, respectively, followed by Section 5 with conclusions.

## 2. Physical and mathematical models

### 2.1. Physical model

This study numerically investigates DDI triggered by a localized hotspot, such as a strong spark [11], in gas and coal particle mixtures, as illustrated in Fig. 2. Previous studies have examined the impact of ignition spot size on detonation propagation [12–14]. Specifically, Lee and Ramamurthi [12] found that variations in ignition source size can significantly influence detonation initiation, propagation, and overall behaviour. Furthermore, Gao et al. [13] identified three typical autoignition reaction front modes associated with changes in hotspot size: supersonic reaction front, detonation development, and subsonic reaction front.

This study aims to investigate the influences of particle surface reactions, gas-solid interactions, and heat (mass) transfer in two-phase detonations. Therefore, the following constant hotspot properties are maintained: a radius of 0.01 m, an initial temperature $T_s$ of 2,500 K, and an initial pressure $p_s$ of 100 atm (i.e., $100p_0$). The premixed gas and particles can heat each other through convective heat transfer, depending on their instantaneous temperature difference. When the particle temperature exceeds a certain threshold, the heterogeneous reactions occur over the particle surface with the oxygen in the gas. The initiation energy for detonation is scaled with the energy content in the gas-solid mixtures, i.e., $\bar{E} = E_{hs}/(E_g + E_s)$, where $\bar{E}$ is the normalised initiation energy, $E_{hs}$ is the initial energy, $E_g$ is the gas phase energy, and $E_s$ is the solid phase energy. The detailed energy quantification for the simulation case in Table 1 is provided in section F of the supplementary document. De-volatilised coal particles, comprising 20% (by mass) inert ash and 80% fixed carbon, are uniformly dispersed within the domain. Due to the low devolatilization rate and minimal volatile gas composition, their impact on detonation development is negligible [15–17]. The particle heat capacity and initial density are 710 J/kg/K and 1,500 kg/m$^3$, respectively.

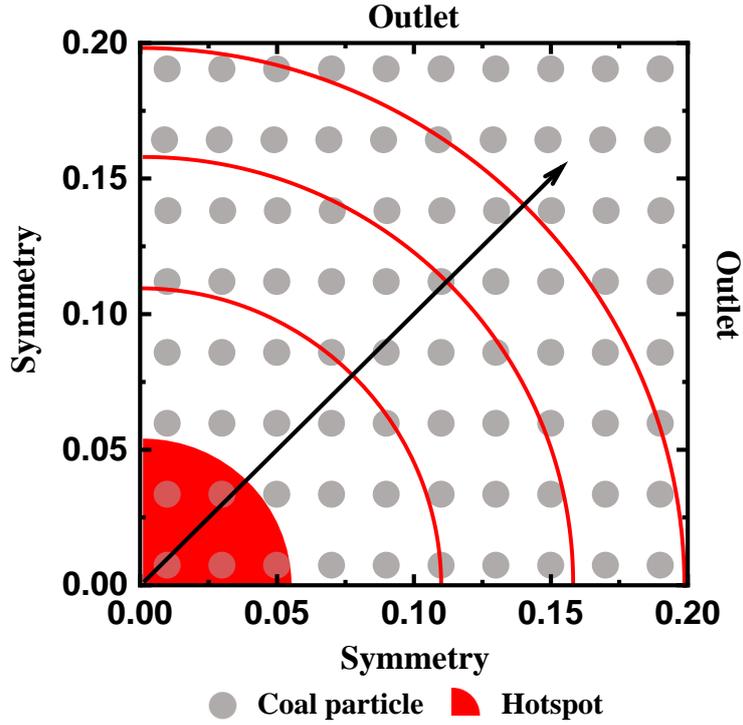

Figure 2: Computational domain and boundary condition. Arrow: detonation propagation direction; red lines: detonation front. Hotspot and particle size are not to scale. Axial label unit: m.

Given the cylindrical geometry, a quarter-section of the domain (0.2×0.2 m$^2$) in Fig. 2 is simulated. The origin of the coordinate system is at the centre of the hotspot, with the *x* and *y* axes aligned with the *symmetry* boundaries. At the two outlets, a *wave-transmissive* condition is enforced for pressure, while a *zero-gradient* condition is applied for other quantities. Numerical soot foils, as depicted in section A of the supplementary document, demonstrate that a uniform Cartesian mesh with grid sizes of 25 and 50 μm yields similar cell regularity and size. Therefore, a 50 μm mesh is utilized in our simulations, resulting in 16 million cells. This configuration ensures a minimum of 194 cells across the half-reaction length (HRL) in the Chapman-Jouguet (C–J) detonation of methane/air mixtures. Detailed resolution analysis is provided in sections B and C of the supplementary document.

In our simulations, the homogeneous gas consists of methane/air mixtures, with an initial temperature of $T_0$ = 500 K and an initial pressure of $p_0$ = 1 atm. Parametric studies are conducted to examine the influence of gas and particle properties on DDI, such as particle concentration, particle diameter, and gas equivalence ratio. The details of the simulated cases are presented in Table 1. A range of parameters will be investigated, including $c_0$ = 10–2,000 g/m$^3$ (Case 1 in Table 1), $d_p$ = 1–30 μm (Case 2), and $\phi$ = 0–4 (Case 3). $\phi$ = 0 represents the homogeneous gas with air only. Additionally, gaseous detonation in a stoichiometric methane/air mixture is included for comparison (Case 0).

Table 1: Summary of simulation conditions

| Case | Particle properties | | Gas properties |
|---|---|---|---|
| | Concentration [g/m$^3$] | Diameter [μm] | Equivalence ratio [-] |
| 0 | 0 | - | 1 |
| 1 | 10 | 1 | 1 |
| | 15 | | |
| | 50 | | |
| | 200 | | |
| | 1,000 | | |
| | 1,750 | | |
| | 2,000 | | |
| 2 | 200 | 1 | 1 |
| | | 1.25 | |
| | | 5 | |
| | | 30 | |
| 3 | 200 | 1 | 0 |
| | | | 0.2 |
| | | | 1 |
| | | | 2 |
| | | | 4 |

**2.2. Governing equations**

The Eulerian–Lagrangian method is used to simulate DDI in two-phase mixtures. The gas and particle equations are solved using the *RYrhoCentralFoam* solver in OpenFOAM-8 [18]. The exchanges of mass, momentum, and energy between the continuous and dispersed phases, including Stokes drag and convective heat transfer, have been validated as detailed in Ref. [19]. Further details are available in section E of the supplementary document. Our numerical approach follows the methodologies used in previous studies [20, 21]; therefore, only key information is presented below.

For the gas phase, combustion of methane/air mixtures is modeled using the GRI 3.0 mechanism, which includes 36 species and 219 reactions (the nitrogen oxide reactions not considered [22]). The effects of coal particles on the gas phase are modelled by the source terms in the equations of mass, momentum, and energy. For the particle phase, the parcel concept [23] is used, where each parcel denotes a collection of particles with identical properties [24, 25]. Initially, these parcels are uniformly distributed over the domain (see Fig. 2). Approximately one million parcels are used, with the number of actual particles per parcel (i.e., particle resolution) determined by the particle size and concentration. Since the particle Biot number is generally small, uniform particle internal temperature is assumed [21]. The surface reaction (or heterogeneous reaction) between the fixed carbon $C_{(S)}$ and oxygen generates carbon dioxide, i.e., $C_{(S)} + O_2 \rightarrow CO_2$. Radiation heat transfer is excluded in our simulations. The equations of particle mass, momentum, and energy respectively read

$$\frac{dm_p}{dt} = -\dot{m}_p, \tag{1}$$

$$\frac{d\mathbf{u}_p}{dt} = \frac{\mathbf{F}_d + \mathbf{F}_p}{m_p}, \tag{2}$$

$$c_{p,p}\frac{dT_p}{dt} = \frac{\beta \dot{Q}_s + \dot{Q}_c}{m_p}, \tag{3}$$

where $t$ is time and $m_p = \pi \rho_p d_p^3/6$ is the particle mass. $\rho_p$ and $d_p$ are the particle material density and diameter, respectively. $\mathbf{u}_p$ is the particle velocity, $c_{p,p}$ the particle heat capacity, $T_p$ the particle temperature, and $\beta$ the fraction of the heat release for particles.

The surface reaction rate is estimated using the kinetic/diffusion-limited rate model, which is based on well-established theories and has been shown to reasonably model coal combustion under various conditions [21, 26, 27]. The validation of this model can be referred to section D of the supplementary document. The surface reaction rate is

$$\dot{m}_p = A_p p_{ox} \frac{D_0 R_k}{D_0 + R_k}, \tag{4}$$

where $A_p$ is the particle surface area and $p_{ox}$ is the partial pressure of the oxidant species in the surrounding gas. The diffusion rate coefficient $D_o$ and kinetic rate coefficient $R_k$ are respectively

$$D_0 = C_1 [\frac{(T + T_p)}{2}]^{0.75}/d_p \text{ and } R_k = C_2 e^{-(E_a/RT_p)}, \tag{5}$$

where the constants $C_1$ and $C_2$ are $5.06 \times 10^{-7}$ s/K$^{0.75}$ and 0.006 s/m, whilst the activation energy

$E_a$ is 50 kJ/mol [28-30]. $T$ is the gas temperature, and $R$ is the universal gas constant.

In Eq. (2), the Stokes drag is $\mathbf{F}_d = (18\mu/\rho_p d_p{}^2)(C_d Re_p/24) m_p (\mathbf{u} - \mathbf{u}_p)$ [31], where $\mu$ is the gas dynamic viscosity and $C_d$ is the drag coefficient, following Schiller and Naumann [32]. $Re_p \equiv \rho d_p |\mathbf{u}_p - \mathbf{u}|/\mu$ is the particle Reynolds number, where $\rho$ and $\mathbf{u}$ are the density and velocity of the gas phase, respectively. $\mathbf{F}_p$ represents the pressure gradient force and is calculated as $\mathbf{F}_p = -V_p \nabla p$, where $V_p$ is the particle volume and $p$ is the pressure. In Eq. (3), $Q_s$ is the heat release rate of the particle surface reaction. $\dot{Q}_c = h_c A_p (T - T_p)$ is the convective heat transfer, where $h_c$ is the convective heat transfer coefficient, estimated with the Ranz and Marshall correlation [33].

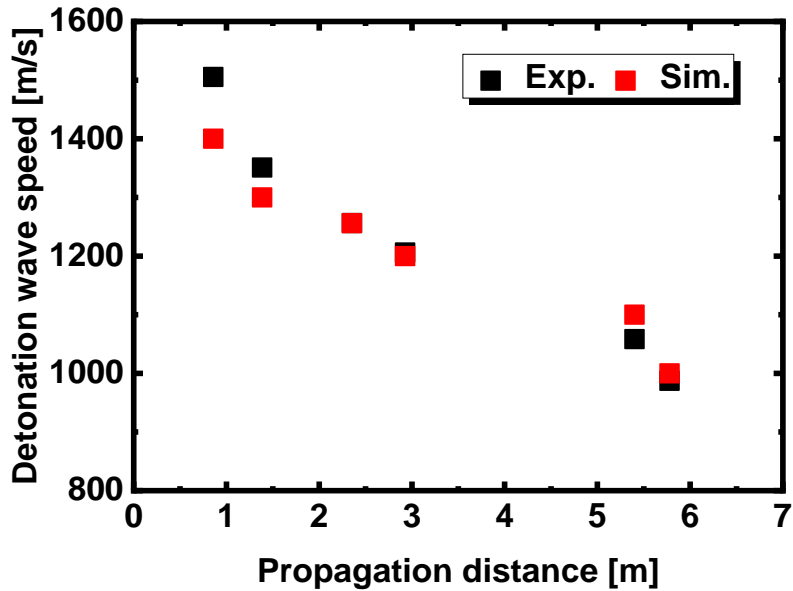

Figure 3: Change of detonation speed with propagation distance. Experimental data: Pinaev and Pinaev [34].

**2.3 Model validation**

Validations of burning rate and ignition delay time in particle ignition problems using the *RYrhoCentralFoam* solver can be found in Refs. [35, 36]. In this study, we further assess the solver's accuracy by reproducing coal particle detonation experiments conducted by Pinaev and Pinaev [34]. In their experiments, the detonation wave speed was measured using a cylindrical explosion tube with a length of 6.75 m and a diameter of 0.07 m. The initial pressure and temperature were approximately 0.1 MPa and 300 K, respectively. The coal particles, comprising 78.4% carbon (by mass) and 21.6% ash, were used. The hotspot energy was 2.6 kJ, resulting in an energy density of approximately 16.5 MJ/m$^2$, based on the Ref. [37], which aligns with the experimental range of 3.8–27.6 MJ/m$^2$ reported in Ref. [34]. A comparison of the simulated and experimental detonation speeds is shown in Fig. 3, demonstrating that our model can reasonably predict the detonation speed in coal particle suspensions, despite uncertainties in coal particle sizes and/or physicochemical properties in the experiments.

## 3. Results
### 3.1 Particle concentration effects

Figure 4 illustrates the evolution of the leading shock front (LSF) speed with particle concentrations ranging from $c_0$ = 0 to 2,000 g/m$^3$. The particle diameter was $d_p$ = 1 μm, and the gas equivalence ratio was unity. The leading shock speed ($D_{LSF}$) was calculated along the diagonal of the domain shown in Fig. 2. A successful DDI is characterised by the consistent coupling of the leading shock front with the reaction front (RF), enabling continuous outward propagation [38–40]. As shown in Fig. 4, successful DDI occurs only at $c_0$ = 50, 200, and 1,000 g/m$^3$. At lower concentrations (e.g., 10 g/m$^3$) or higher concentrations (e.g., 2,000 g/m$^3$), detonations initiated near the hotspot, but the shock intensity decreased rapidly, leading to the eventual decoupling of the LSF and RF. The causes of LSF intensity attenuation differed at $c_0$ = 10 and 2,000 g/m$^3$. At $c_0$ = 10 g/m$^3$, the LSF intensity was reduced due to insufficient particle loading, which limited the energy contribution from burning particles. In contrast, at $c_0$ = 2,000 g/m$^3$, the LSF was significantly weakened by denser particle suspensions, resulting in extensive momentum extraction [41].

In successful detonations ($c_0$ = 50, 200, and 1,000 g/m$^3$), the average LSF speed initially increased before decreasing with particle concentration, peaking at approximately 1,600 m/s for $c_0$ = 200 g/m$^3$. At $c_0$ = 50 g/m$^3$, reduced heat release from surface reactions resulted in a lower degree of LSF intensification. At $c_0$ = 1,000 g/m$^3$, the dispersed phase absorbed more heat and momentum. Additionally, the detonation speed in our results oscillates towards the end of the computational domain, indicating that the detonation reaches a stable state after propagating over a long distance.

The detonation speed was significantly lower than $90\%D_{CJ}$ ($D_{CJ} \approx 1,800$ m/s in methane/air mixtures), similar to that of a purely gaseous detonation [42], due to the curvature effect. Specifically, at $c_0 = 1,000$ g/m³, the average speed was only about 45% of $45\%D_{CJ}$, with the LSF speed oscillating between 80 and 120% of the C–J speed at $c_0 = 50$ g/m³. This is due to the combined effects of frontal curvature [43] and dispersed phase [44].

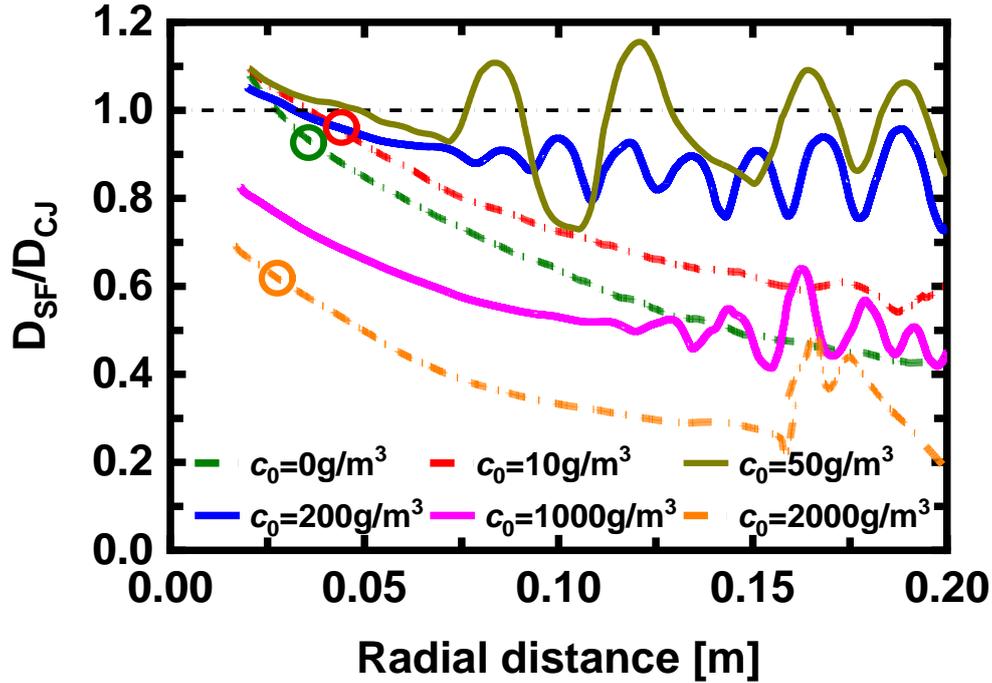

Figure 4: Change of leading shock speed with radial distance with different particle concentrations. Dash-dotted line: CJ speed in methane/air mixtures. Solid line: successful DDI; Dashed line: failed DDI. $d_p = 1$ μm and $\phi = 1$. Circle: detonation decoupling location.

Figure 5 illustrates the peak pressure trajectories for the cases shown in Fig. 4. In all cases, overdriven and triple-point-free detonations initiate near the hotspot. When DDI fails, specifically at $c_0 = 0$, 10, and 2,000 g/m³, distinct peak pressure trajectories are not observed. For $c_0$ values between 50 and 1,000 g/m³, the detonation cells are distinguishable. The cell size is calculated using $\lambda = (l_1 + l_2)/2$, where $l_1$ and $l_2$ denote the two diagonal lines of the detonation cell in Fig. 5(c). Notably, the characteristic length is closely related to the cell size [45, 46]. In Fig. 5, the average cell size initially decreases before increasing as particle concentration rises: for $c_0 = 50$, 200, and 1,000 g/m³, the measured cell sizes are approximately 18, 6, and 9 mm, respectively. This is attributed to the dual competitive effects of the coal particles on detonative combustion: heat absorption and combustion heat release. At relatively low particle concentrations, the heat from surface reactions is insufficient, resulting in longer induction lengths and affecting interactions between burning particles and the LSF. In contrast, at very high particle concentrations, the heat absorption effect becomes pronounced due to the limited availability of oxygen, significantly weakening the detonation. Consequently, detonation intensity is reduced at both extremely low and high particle concentrations. In Fig. 5(f), the DDI fails at $c_0 = 2,000$ g/m$^3$, where shock focusing, as highlighted by Xu et al. [47], occurs at $R \approx 0.16$ m. However, the focal pressure and temperature are insufficient to trigger a new detonative spot, underscoring the challenges in achieving successful DDI at extreme particle concentrations.

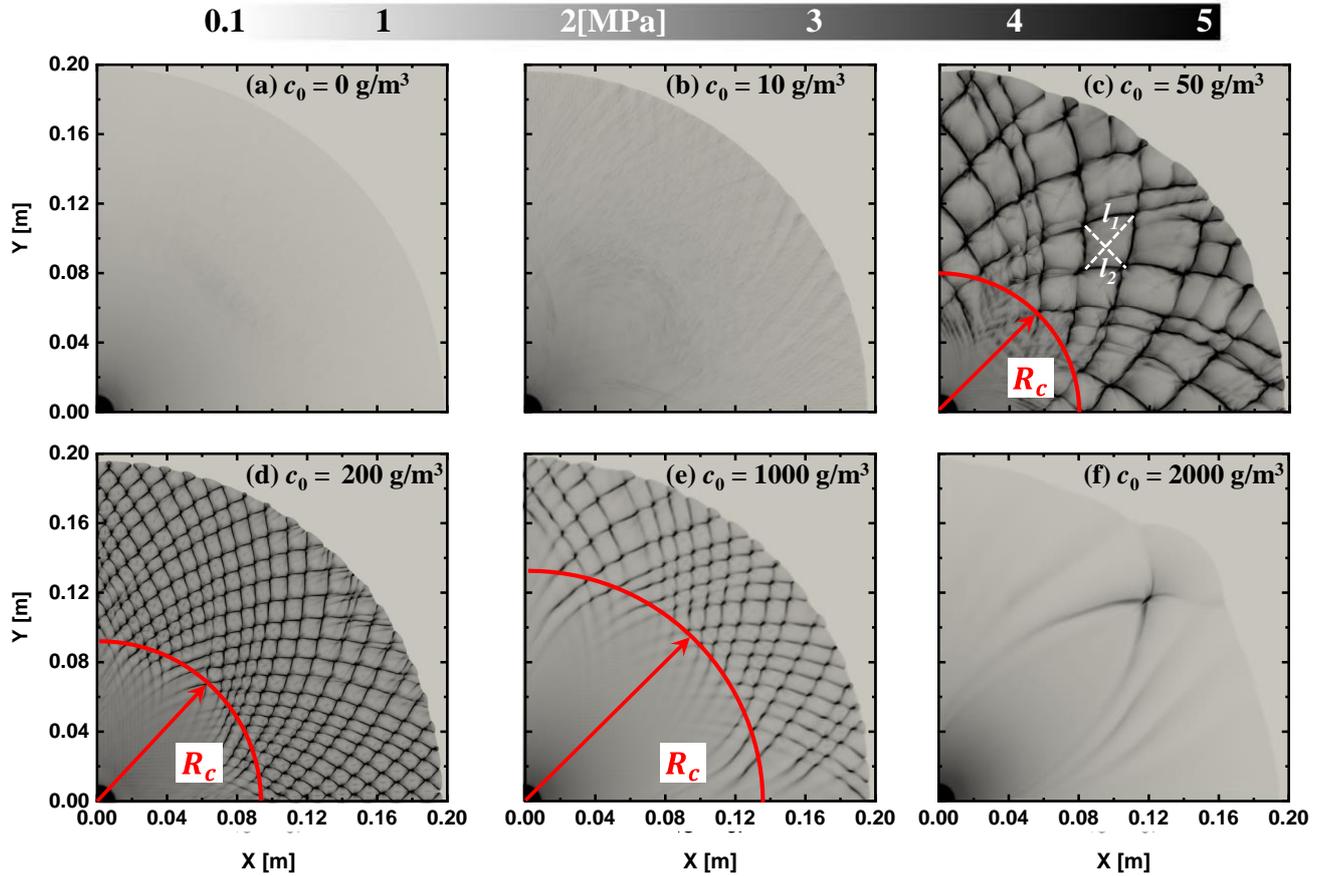

Figure 5: Trajectories of peak pressure with different particle concentrations. $l_1, l_2$: diagonal distance of a detonation cell. $R_c$: cellular detonation radius. $d_p = 1$ μm and $\phi = 1$.

Figure 6 depicts the time history of average maximum pressures ($\bar{p}_{max}$) within the domain. From this, the detonation development process can be divided into three stages: overdriven detonation (Stage I), developing cellular detonation (Stage II), and cellular detonation (Stage III).

During Stage I, the peak pressure rapidly decays to a low value (e.g., approximately 2.5 MPa at $c_0$ = 50 g/m³ in Fig. 6a) due to the absence of triple points. In Stage II, transverse waves begin to emerge, and the front progressively becomes cellular. The appearance of triple points leads to an increase in peak pressure and larger pressure fluctuations. Stage III is characterised by a stable propagating cellular detonation, with $\bar{p}_{max}$ values of 9.3, 15.6, and 13.6 MPa for $c_0$ = 50, 200, and 1,000 g/m³, respectively.

Both the frequency and amplitude of pressure oscillations are crucial indicators for assessing explosion risks [48]. In Stage III, spectrum analysis is used to determine the dominant frequency of pressure oscillations in Fig. 6. The results show that both the maximum pressure oscillation amplitude and frequency initially increase, then decrease, as particle concentration rises. The durations of Stage I ($\tau_1$) and Stage II ($\tau_2$) at various particle concentrations are compared in section K of the supplementary document, revealing that both $\tau_1$ and $\tau_2$ increase with higher particle concentrations. This trend is expected, as particle heating by the surrounding gas slows down with increasing particle concentration.

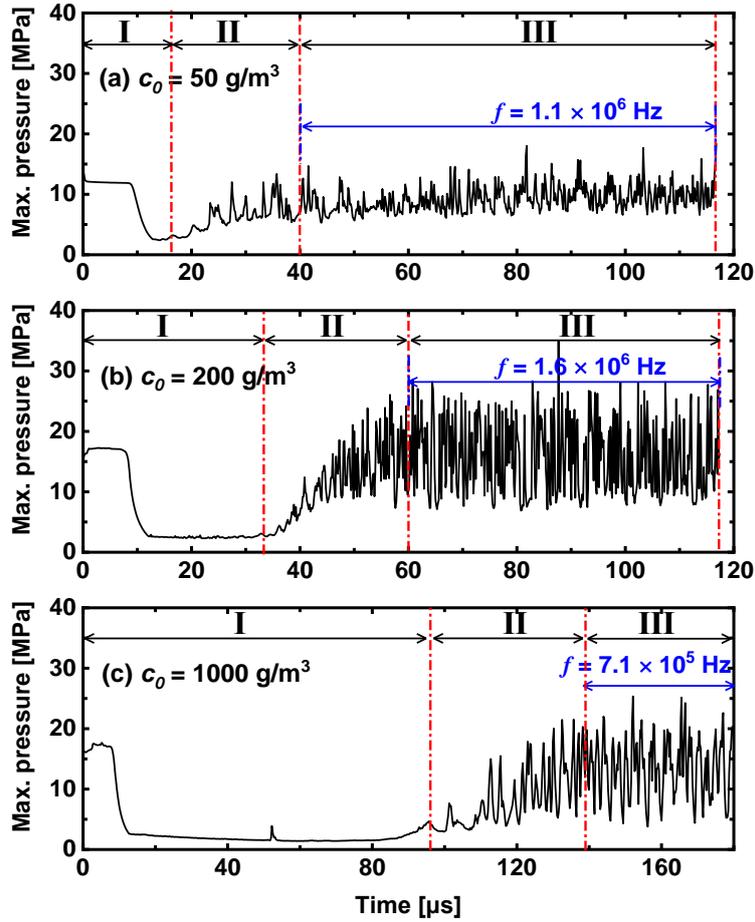

Figure 6: Temporal evolution of average maximum pressures for $c_0$ = 50, 200 and 1,000 g/m$^3$. Stage I: Overdriven detonation; Stage II: Developing cellular detonation; Stage III: Cellular detonation. $d_p$ = 1 μm and $\phi$ = 1.

The cellular detonation radius, $R_c$, is defined as the distance from the point where detonation cell trajectories become distinct (transverse waves are enhanced) to the origin of the domain, marking the onset of Stage III. $R_c$ is a crucial parameter for overpressure regulation in explosion prevention and prediction [49]. As particle concentration increases from 50 to 1,000 g/m$^3$, $R_c$ increases monotonically, as shown in Fig. 5. Cell formation, accompanied by the evolution of transverse waves, occurs first at $R_c \approx 0.055$ m for $c_0$ = 50 g/m$^3$. For $c_0$ = 200 g/m$^3$, cells form at $R_c \approx 0.092$ m. However, at $c_0$ = 1,000 g/m$^3$, cell formation is significantly delayed, with $R_c \approx 0.14$ m.

### 3.2 Particle diameter effects

Figure 7 illustrates the evolution of LSF speed across a range of particle diameters from 1 to 30

μm. The particle concentration is $c_0 = 200$ g/m$^3$, and the gas equivalence ratio is $\phi = 1$. In all cases, overdriven detonations initiate immediately beyond the hotspot. However, particle size significantly affects the subsequent propagation. Specifically, for particle diameters of 5 and 30 μm, the LSF speed gradually decreases to only 40–55% of the C-J speed of the gaseous mixtures as the detonation wave propagates outward, leading to DDI failure. With finer particles, such as $d_p = 1$ and 1.25 μm, successful detonation transmission is observed in the particle suspensions, with average speeds lower ($\approx 80\% D_{CJ}$) than the C–J speed. Notably, particle-free mixtures (i.e., methane/air mixtures) cannot be detonated under these conditions, indicating that micro-sized particles play a crucial role in facilitating DDI. Furthermore, the detonation speed fluctuations for $d_p = 1$ μm are more pronounced than those for $d_p = 1.25$ μm, despite the slight difference in diameter. This suggests that detonation speed is highly sensitive to the size of fine particles, and therefore, special consideration should be given to particles or dusts in this size range for effective explosion mitigation.

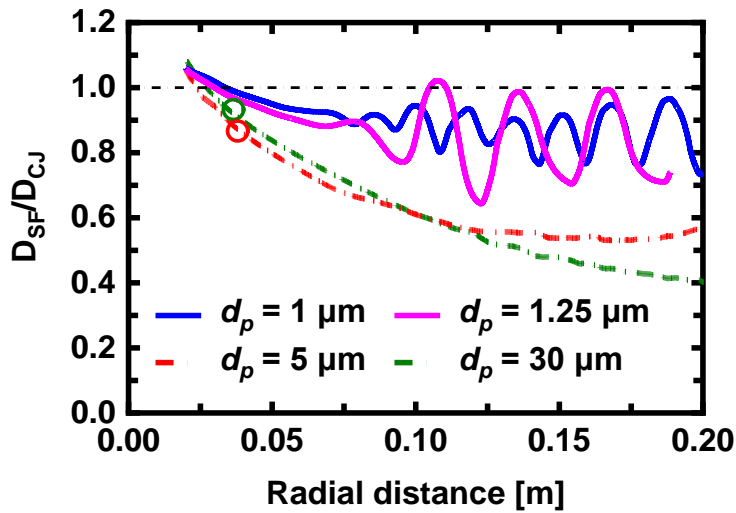

Figure 7: Change of leading shock speed with radial distance with different particle diameters. Dash-dotted line: the CJ speed in methane/air mixtures; Solid line: successful DDI; Dashed line: failed DDI. $c_0 = 200$ g/m$^3$ and $\phi = 1$. Circle: detonation decoupling location.

Figure 8 illustrates the peak pressure trajectories for $d_p = 1$ and 1.25 μm, corresponding to the conditions in Fig. 7. The average cell sizes for the two diameters are 0.06 and 0.012 m, respectively. The results show that even slightly larger particles lead to increased cell sizes. This is because smaller particles generally release more heat [50], effectively reducing the distance between the LSF and RF, thereby decreasing the cell structure. Consistent with the findings in Ref. [51], the results also indicate that particle size affects frontal stability. Additionally, when $d_p = 1$ μm, the development of secondary cells occurs, initially featuring weak transverse waves that intensify as the detonation propagates. This eventually results in the formation of multiple cells within the primary cell structure, as shown in the white box in Fig. 8(a).

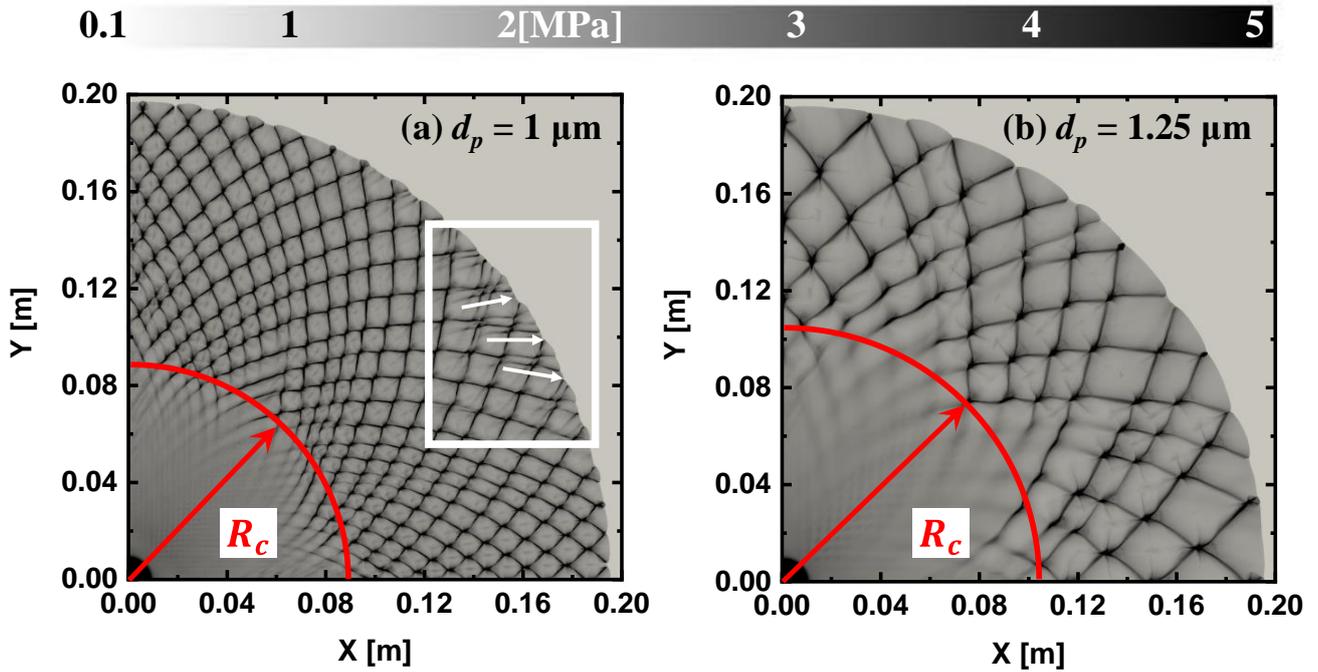

Figure 8: Trajectories of peak pressure with $d_p = 1$ and 1.25 μm. The white arrows indicate the transverse waves in primary cells. $c_0 = 200$ g/m$^3$ and $\phi = 1$.

The coal particle size can significantly influence momentum and energy transfer. Specifically, smaller coal particles facilitate higher rates of momentum and energy transfer. The rate of momentum exchange depends on the ability of coal particles to respond quickly to changes in gas flow. Smaller particles can adjust their speed more rapidly, enhancing momentum exchange. Additionally, these smaller particles have a higher specific surface area, which accelerates momentum transfer. Figure 9 shows the temporal evolution of the momentum transfer rate along the diagonal direction of the domain, indicating that smaller particles, such as $d_p = 1$ μm, result in a more significant momentum exchange. For a given particle concentration, the specific surface area (S) is inversely proportional to the particle diameter ($d_p$) (i.e., $S \propto 1/d_p$). This means that larger particles have less surface area available for interactions with the gas phase, which is crucial for heat transfer. As the particle diameter increases, the total surface area decreases, thereby reducing the rate of heat absorption from the gas to the particles. Additionally, energy transfer from the particle surface to the homogeneous mixture is described by $Q = (T_p - T)/\tau_{pg}$, where $\tau_{pg}$ is the characteristic time for energy transfer, $T_p$ is the particle temperature, and $T$ is the gas temperature [44]. As the particle diameter increases, the characteristic time for heat transfer also increases, being inversely proportional to the particle diameter. Figure 9(b) shows the spatial evolution of the energy transfer rate along the diagonal direction, demonstrating that heat transfer between the two phases decreases as the particle diameter increases.

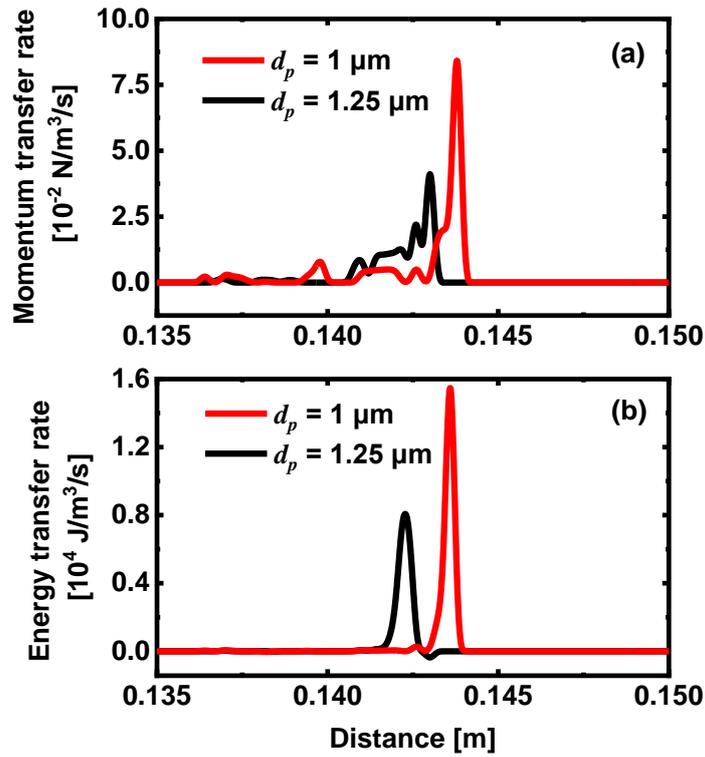

Figure 9: Spatial evolution of momentum and energy transfer rates along the diagonal direction.

Figure 10 shows the average maximum pressures for different particle sizes. In Stage III, the average maximum pressures $\bar{p}_{max}$ are 15.6 and 9.2 MPa for $d_p = 1$ and 1.25 μm, respectively. Higher oscillation frequencies are observed for $d_p = 1$ μm, due to the increased specific surface area of the particles, which promotes more vigorous heterogeneous reactions. Additionally, a larger particle diameter corresponds to a larger radius of cellular detonation $R_c$. Additionally, both $\tau_1$ and $\tau_2$ increase with larger particle diameters, as detailed in section K of the supplementary document.

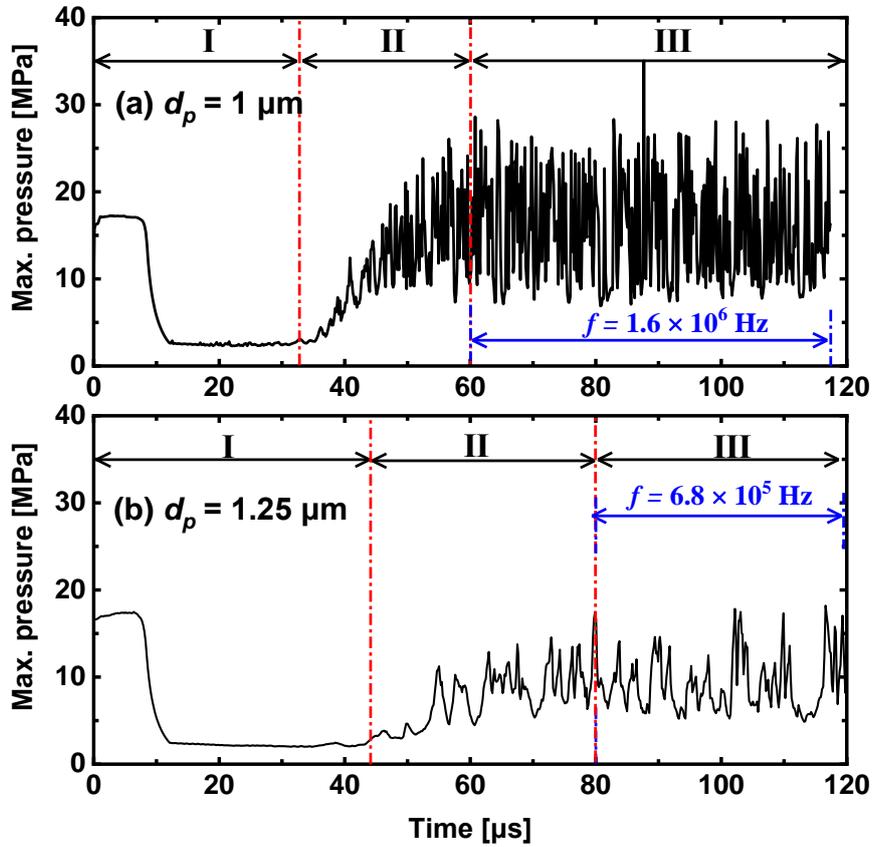

Figure 10: Temporal evolution of average maximum pressures for $d_p$ = 1 and 1.25 μm. $c_0$ = 200 g/m$^3$ and $\phi$ = 1.

### 3.3 Gas equivalence ratio effects

Figure 11 shows the evolution of the LSF speed for different gas equivalence ratios. The particle diameter is 1 μm, and the concentration is $c_0$ = 200 g/m$^3$. Detonation initiation and propagation are successfully achieved for $\phi \leq 2$. As $\phi$ increases, the average LSF speed gradually decreases. A smaller $\phi$ leads to earlier oscillations of the LSF speed, indicating a more rapid onset of wave cellularisation. When $\phi$ reaches 4, the detonation decouples at $R \approx 0.03$ m. This is due to the increase in $\phi$, which reduces the available oxygen for particle surface reactions. Consequently, the energy from surface reactions becomes insufficient to sustain the detonation structure, resulting in a failed DDI.

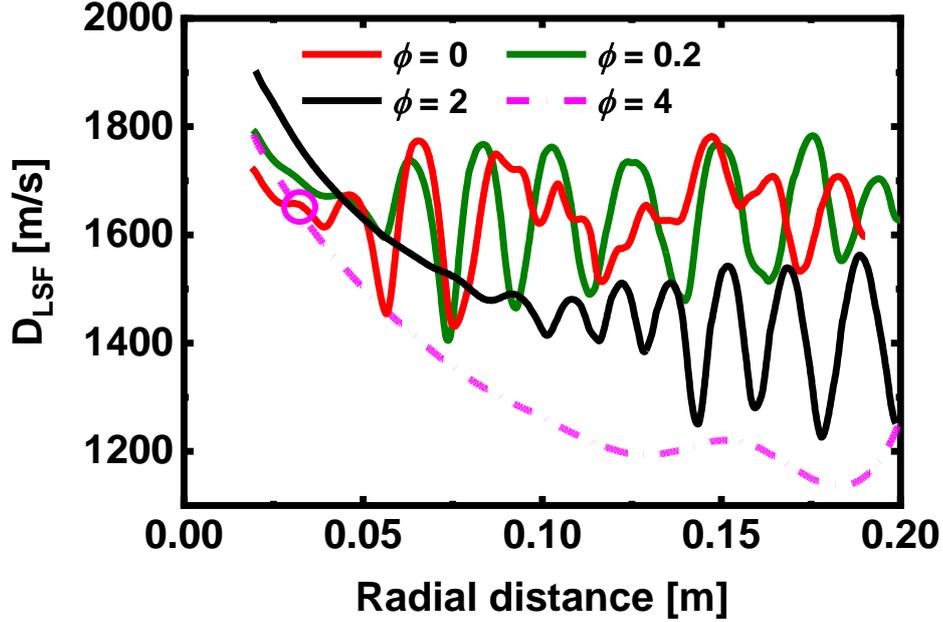

Figure 11: Change of leading shock speed with radial distance with gas equivalence ratio from 0 to 4. Solid line: successful DDI; Dashed line: failed DDI. $c_0$ = 200 g/m$^3$ and $d_p$ = 1 μm. Circle: detonation decoupling location.

Different detonation cellular structures are observed in Fig. 12 at varying gas equivalence ratios. The cellular detonation radius, $R_c$, gradually increases as $\phi$ increases, with the smallest radius ($R_c \approx$ 0.05 m) achieved at $\phi$ = 0 (i.e., the methane-free case). The average cell size remains relatively consistent ($\approx$ 0.006 m) for $\phi$ = 0–4. In contrast to the results from gaseous detonation experiments in stoichiometric methane/oxygen/nitrogen mixtures [52], where cell sizes typically decrease and then increase with changes in gas equivalence ratios, the results here show that when $\phi$ = 1 (see Fig. 5d), the gas reactions are relatively intense, leading to more uniform cell distributions. This indicates that gas reactions contribute to the stability of detonation propagation in two-phase detonations, although they are not the sole determinants of DDI success. Additionally, multiple occurrences of kinked fronts caused by the curvature of cylindrical detonations [53] are highlighted in Figs. 12(a)–(b).

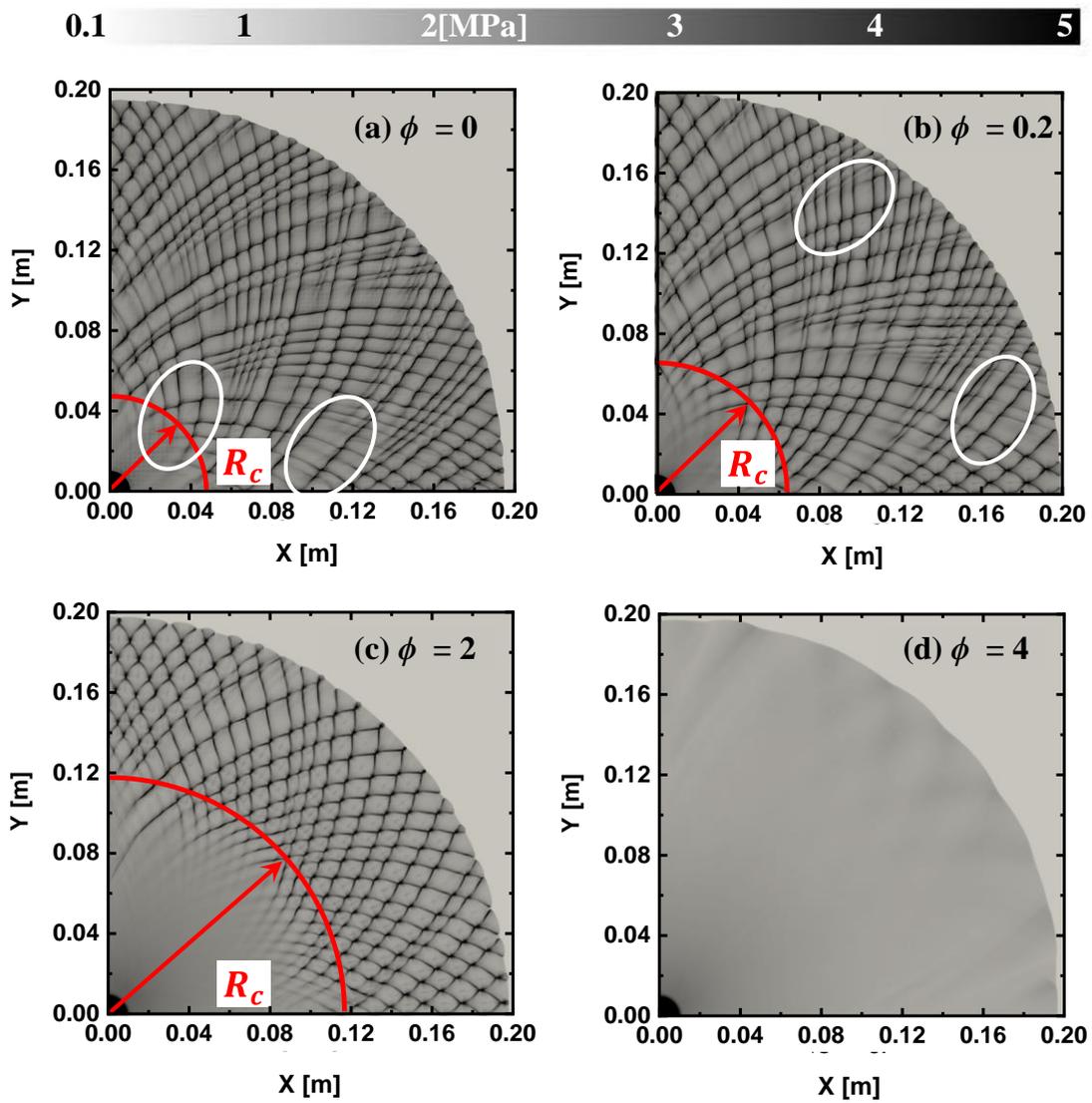

Figure 12: Trajectories of peak pressure with gas equivalence ratio from 0 to 4. $c_0$ = 200 g/m³ and $d_p$ = 1 μm.

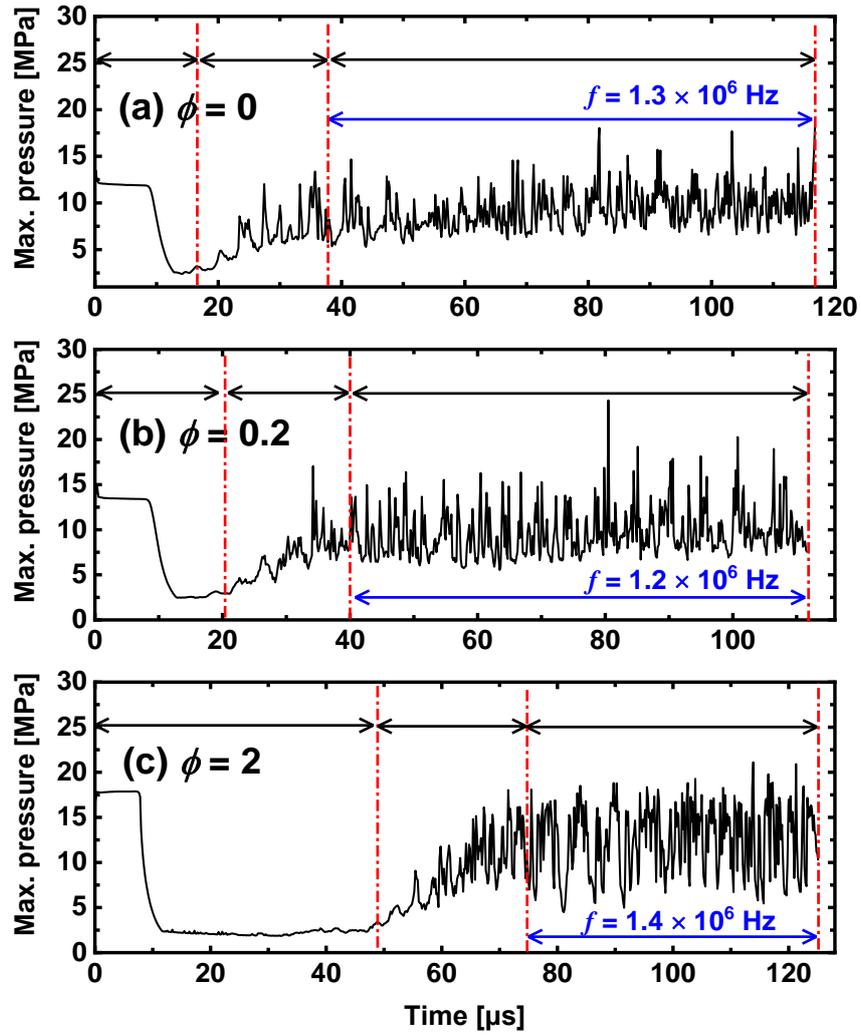

Figure 13: Temporal evolution of average maximum pressures for various gas equivalence ratio. $c_0 = 200$ g/m$^3$ and $d_p = 1$ μm.

As observed in Fig. 13, $\bar{p}_{max}$ during Stage III for $\phi = 0, 0.2$, and 2 are 9.3, 9.6, and 11.4 MPa, respectively. It initially increases and then decreases, peaking at $\phi = 1$ (≈ 15.6 MPa, as shown in Fig. 6(b)). The maximum amplitudes and oscillation frequencies of $\bar{p}_{max}$ exhibit a similar non-monotonic trend as $\phi$ increases, mirroring the observations in purely gaseous detonations [54]. Notably, the minimum pressure amplitude occurs at $\phi = 0$, indicating that detonations in two-phase mixtures exhibit greater intensity compared to those in purely particulate detonations [55]. Furthermore, the prolongation of Stages I and II, reflected in increased $\tau_1$ and $\tau_2$ (see section K of the supplement document), is due to the weakening of coal particle surface reactions.

## 4. Discussion
### 4.1 Detonation structure

The detonation structure in two-phase mixtures is shown in Fig. 14. The particle conditions are $c_0$ = 50 g/m³ and $d_p$ = 1 μm, with a gas equivalence ratio of $\phi$ = 1. In Figs. 14(a) and 14(b), the incident wave (IW), Mach stem (MS), triple point (TP), transverse wave (TW), and LSF are clearly captured. Based on the characteristics of the particle distribution, Figs. 14(d)–(f) categorise the dispersed particles into three types: inert ash (S1 region, burned particles), burning particles (S2 region), and unburned particles (S3 region, ahead of the shock front, SF), as marked in Fig. 14(d). In the S1 region (R ⩽ 0.06 m), the fixed carbon is almost completely consumed, leaving primarily ash. In comparison to the regions of $R \leq 0.06$ m, the particles are denser within the S2 region, indicating a non-uniform distribution of coal particles. This non-uniformity occurs as the coal particles move with the detonation wave, gradually accumulating behind it. Furthermore, in the S2 region, the combustion progress of individual carbon particles varies, as illustrated by the particle mass ($m_p$) and the mass fraction of the fixed carbon ($Y_{C(S)}$) in Figs. 14(e)–(f). High concentrations of incomplete combustion products, such as CO, can be observed in the S2 region. Figure 14(c) shows that the CO mass fraction ($Y_{CO}$) is lower in the S1 region than in the S2 region, while the mass fraction of complete combustion products, such as H₂O, exhibits the opposite trend. Specifically, the H₂O mass fraction ($Y_{H_2O}$) is higher in the S1 region than the S2 region. Detailed analysis can be found from sections G and H of the supplementary document. This is due to the accumulation of particles behind the detonation wave, where the particles compete with methane for oxygen, resulting in incomplete combustion of methane. In the S3 region, the detonation wave has not yet reached, and therefore all variables remain in their initial states.

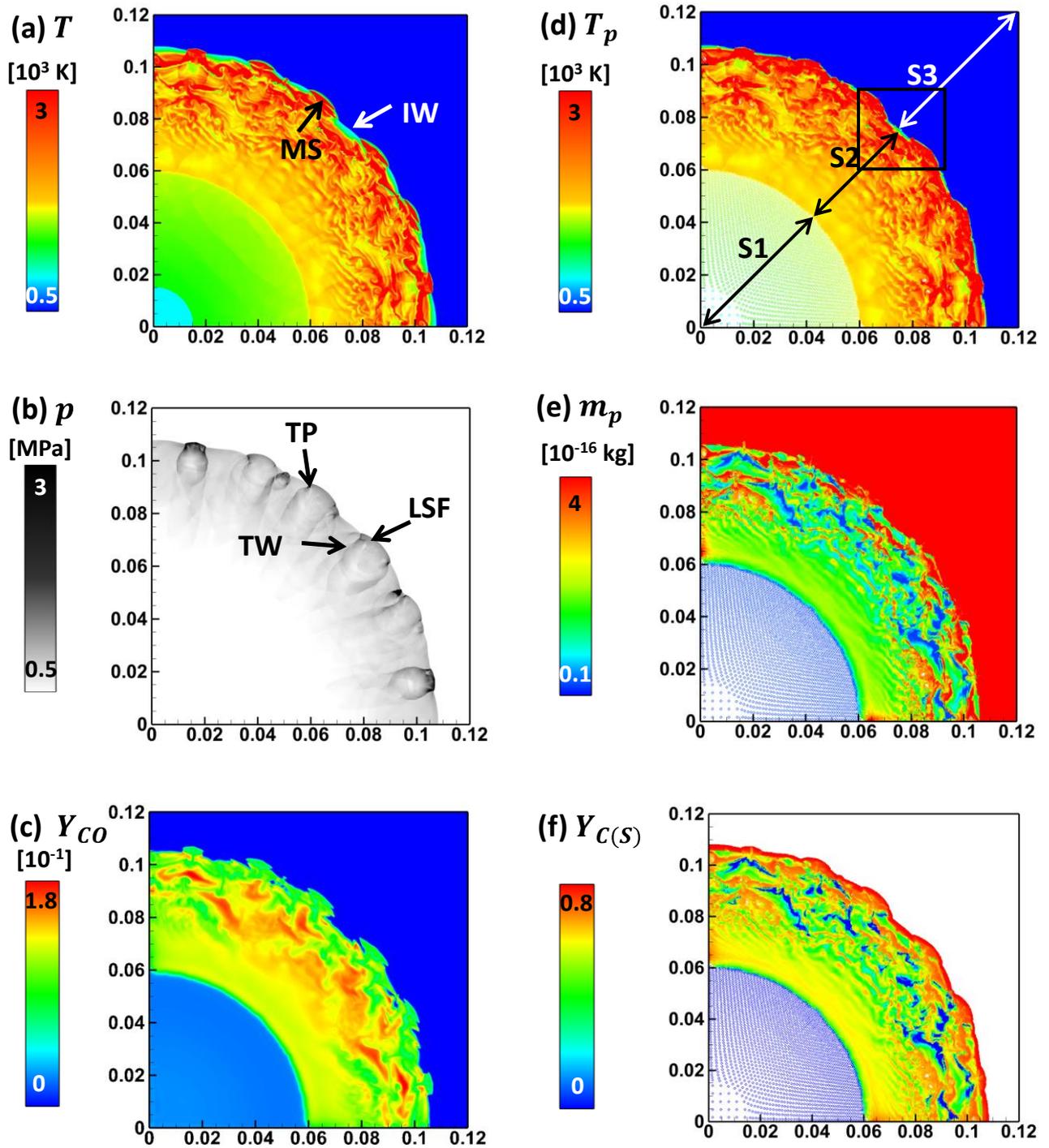

Figure 14: Contours of (a) pressure, (b) gas temperature, (c) mass fraction of the carbon monoxide, (d) particle temperature, (e) particle mass, (f) mass fraction of the fixed carbon. LSF: leading shock front. MS: Mach stem. IW: incident wave; TW: transverse wave. TP: triple point. Axial label unit: m. $c_0$ = 50 g/m$^3$, $d_p$ = 1 μm, and $\phi$ = 1.

Figure 15 shows the distribution of coal particles behind the LSF, coloured by particle temperatures ($T_p$). Before the LSF, particle temperatures remain at their initial levels and are therefore not visualised in the figure. In the induction zone, particles are heated to over 2,000 K by both the LSF and RF, promoting rapid surface reactions that supply substantial energy for detonation propagation.

Two reaction fronts are observed in Fig. 15: the gas reaction front (RFg) and the solid reaction front (RFs). Across the incident wave and Mach stem, the gas reaction front consistently precedes the solid reaction front, supporting the findings of Azadboni et al. [56] and Zhang et al. [57]. Consequently, the entire process of coal particle detonation can generally be divided into three stages: particle heating, methane combustion, and particle burning.

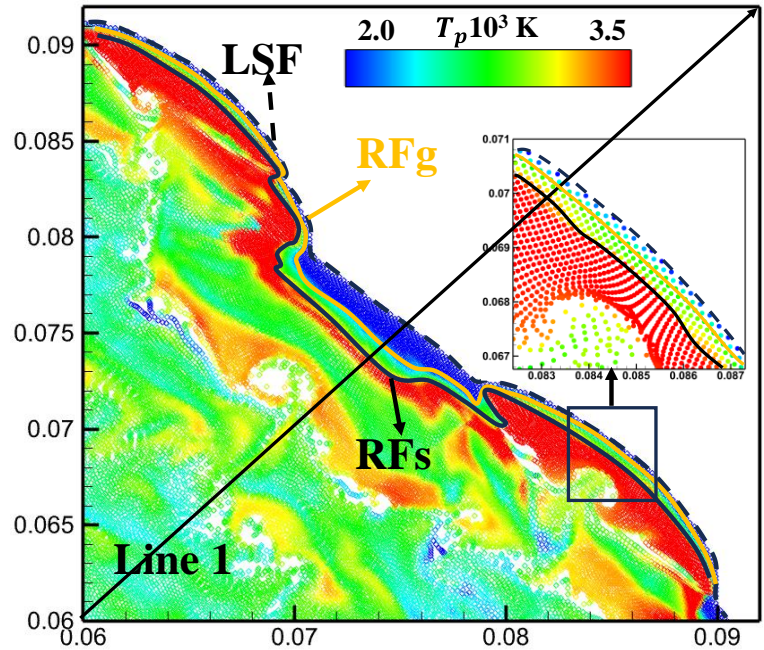

Figure 15: Particle temperature around the detonation front. RFs: solid reaction front. RFg: gas reaction front. LSF: leading shock front. IZ: induction zone. Particles are not shown before LSF.

Figure 16 illustrates the distribution of coal particles at three instants. In region D, prior to the arrival of the detonation wave (at 55 μs in Fig. 16a), the particles are uniformly distributed. As the detonation progresses to 56 μs, the particle distribution becomes denser. By 57 μs, as the wave continues to propagate, the distribution appears sparser. The evolution of coal particles can be attributed to two main factors. Firstly, wave disturbances significantly influence particle distribution. Particles tend to cluster densely near transverse waves, while regions adjacent to rarefaction waves exhibit sparser distributions. Secondly, the non-uniform flow velocity leads to the formation of shear layers, which eventually develop into vortices [58]. These vortices play a crucial role in reshaping the local flow structures, thereby affecting the spatial distribution of particles. High-vorticity regions drive the particles, creating locally sparse areas, while low-vorticity regions encourage particle accumulation.

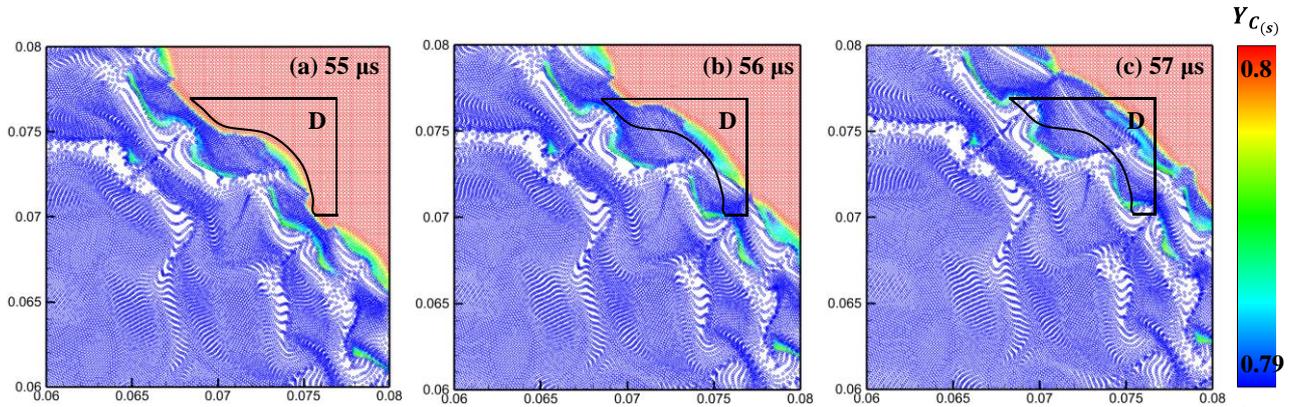

Figure 16: Distribution of particles around the detonation wave at different instants. $c_0 = 50$ g/m$^3$, $d_p$ = 1 μm, and $\phi = 1$. Axial label unit: m.

Figure 17 shows the propagation of the blast wave outward from the hotspot over a period of 2 μs to 60 μs in the coal particle/air mixtures, with $\phi = 0$. As shown in Fig. 17(a), as the shock wave progresses, a high concentration of particles accumulates in the central section of the domain, forming an annular combustion zone (M). The inner and outer boundaries of this zone are denoted by SL1 and SL2, respectively. In Figs. 17(m)–(p), the mass fraction of CO ($Y_{CO}$) is elevated, as CO produced by surface reactions can further react (CO$_2$ → CO) in the gas environment, increasing the production of toxic gas. As the detonation advances, by 15 μs, the annular combustion zone expands, and local concave surfaces form, as seen in Fig. 17(b). In the coal particle/air mixtures, these local concave flame surfaces develop into hotspots, forming the characteristic structures of triple points and the Mach stem. The newly generated hotspots continue to collide, ensuring the complete success of detonation initiation, which then propagates steadily (refer to section I of the supplementary document).

By 30 μs, the detonation cells begin to emerge, with the folding of the inner boundary intensifying. As shown in the enlarged section of Fig. 17(c), forward and backward jets formed by the collision of transverse waves (TWs) with burned materials are observed, as noted by Mahmoudi et al. [59]. These jets induce weak perturbations, causing the particle distribution to alternate between dense and dilute areas. Furthermore, the premature consumption of fixed carbon in the dilute regions leads to the formation of a low-temperature zone, highlighted in the enlarged section of Fig. 17(g). As indicated in Fig. 17(d), by 60 μs, the post-detonation flow fully stretches the Rayleigh–Taylor (RT) fingers along the inner boundary (SL1) of the annular combustion zone due to perturbation influences at the interface between burning and burned particles, as mentioned by Tavares et al. [60].

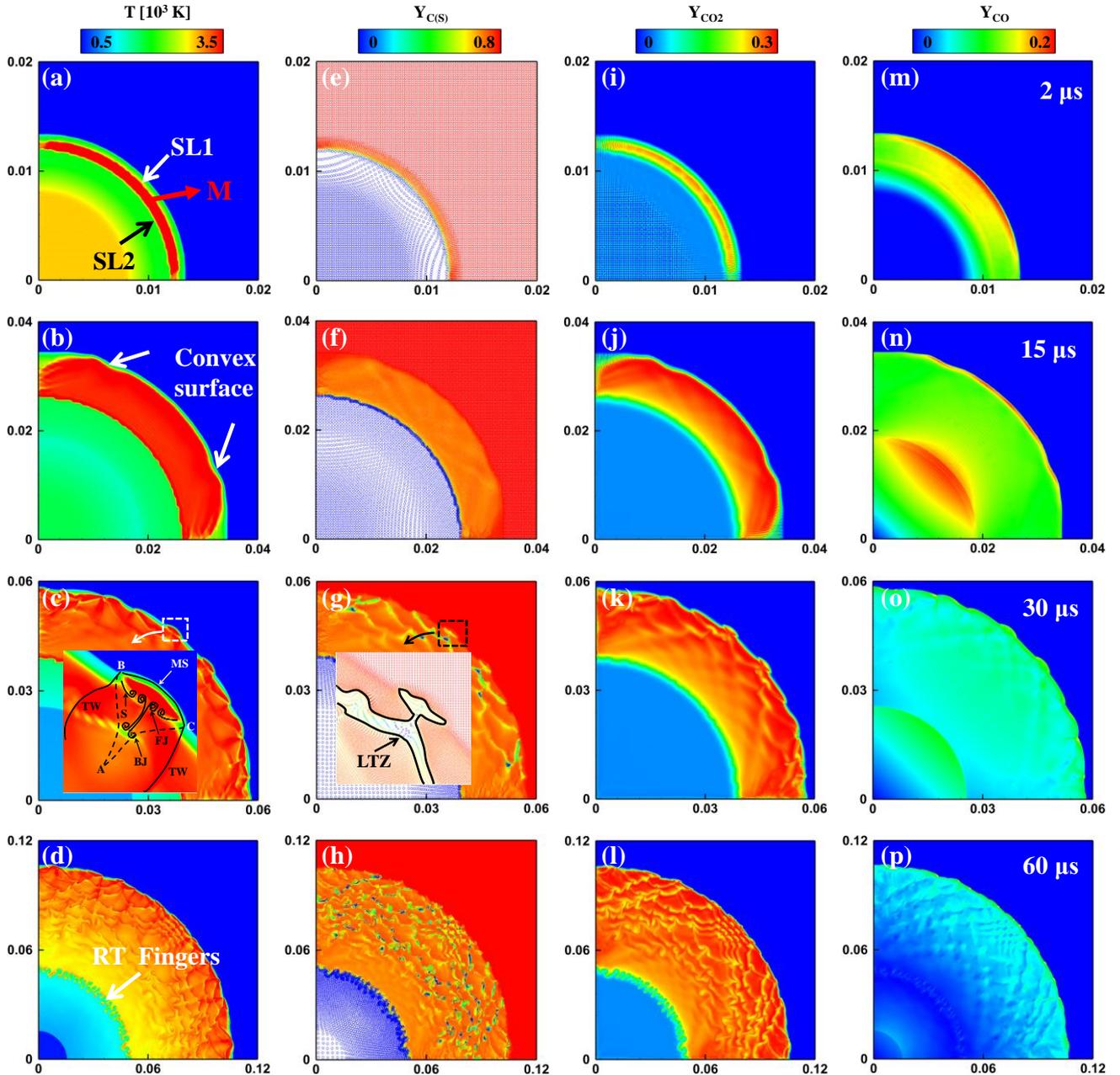

Figure 17: Detonation initiation in coal particles/air mixtures ($\phi = 0$). Left to right the columns show temperature, mass fraction of fixed carbon ($Y_{C_{(S)}}$), carbon dioxide ($Y_{CO_2}$), and carbon monoxide ($Y_{CO}$) at selected times. $c_0 = 200$ g/m³. $d_p = 1$ μm and $\phi = 0$. The enlarged figures: the first half of the cell from A to BC; SL1: inner combustion boundary; SL2: outer combustion boundary; M: annular combustion zone; S: Shear layer; M: Mach steam; TW: Transverse wave; FJ: forward jet; BJ: backward jet; LTZ: low-temperature zone; RT: Rayleigh–Taylor.

**4.2 Direct detonation initiation and propagation**

In Section 3.1, subcritical detonation in methane/air mixtures without reactive particle suspensions is observed in Figs. 4 and 5. However, the introduction of particles significantly alters the DDI process. Figures 18(a)–(c) show the evolution of pressure distribution for three concentrations: $c_0$ = 15, 200, and 1,750 g/m$^3$, corresponding to critical, stable, and cell-free detonations, respectively. Additionally, the LSF speeds, temperature profiles, and peak pressure trajectories for these cases are presented in Figs. 19–20. In the dilute particle case ($c_0$ = 15 g/m$^3$) shown in Fig. 18(a), the peak pressure initially decreases, then abruptly increases, before decreasing again. The DDI process can be divided into four phases based on the pressure profiles: coupling, decoupling, coupling, and decoupling. Figure 19 shows that decoupling ceases once the blast wave decelerates to approximately 1,250 m/s with $c_0$ = 15 g/m$^3$. After this point, the LSF abruptly reaccelerates, resulting in an overdriven detonation. In the diagonal direction, as depicted in Figs. 20(a)–(b), the LSF and RF initially decouple as the blast wave expands, causing the RF to recede from the LSF. Once the LSF and RF are fully coupled, the overdriven detonation wave decays, leading to complete decoupling of the RF, as shown in Figs. 20(c)–(d). Moreover, Figs. 20(e)–(h) illustrate that the cell distribution is irregular and indistinct. Under dilute particle conditions, the fixed carbon is completely consumed, providing energy for the DDI. Due to the low concentration of particles, only a limited amount of heat is released, creating a critical condition.

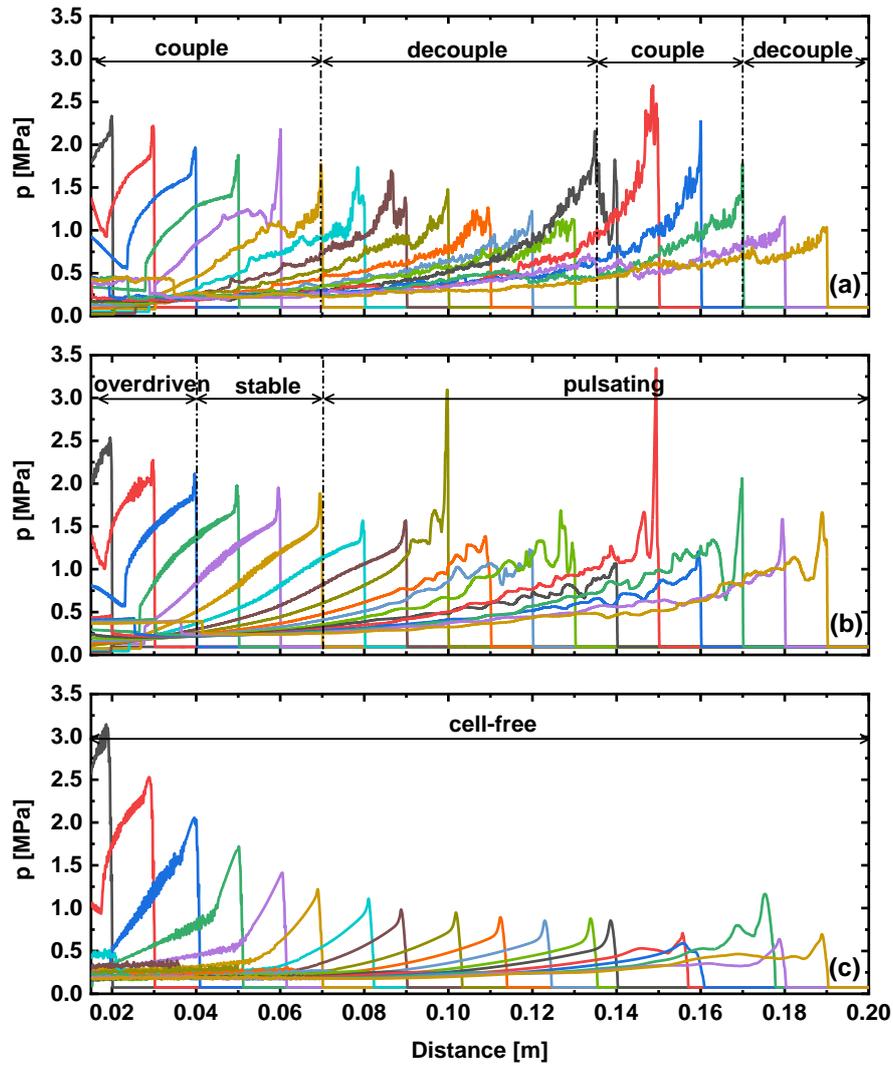

Figure 18: Temporal evolution of pressure distribution along the diagonal direction. Three initiation regimes correspond to (a) critical case with $c_0$ = 15 g/m$^3$, (b) stable case with $c_0$ = 200 g/m$^3$, (c) cell-free case with $c_0$ = 1,750 g/m$^3$. $d_p$ = 1 μm and $\phi$ = 1.

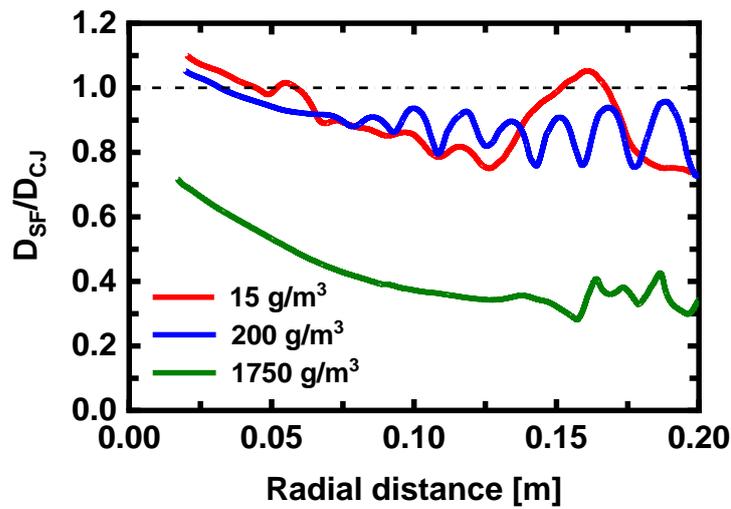

Figure 19: Change of leading shock speed with radial distance with various particle concentrations. The dashed line corresponds to the CJ detonation speed in methane/air mixtures, with $D_{CJ}$ = 1,800 m/s. $d_p$ = 1 μm and $\phi$ = 1.

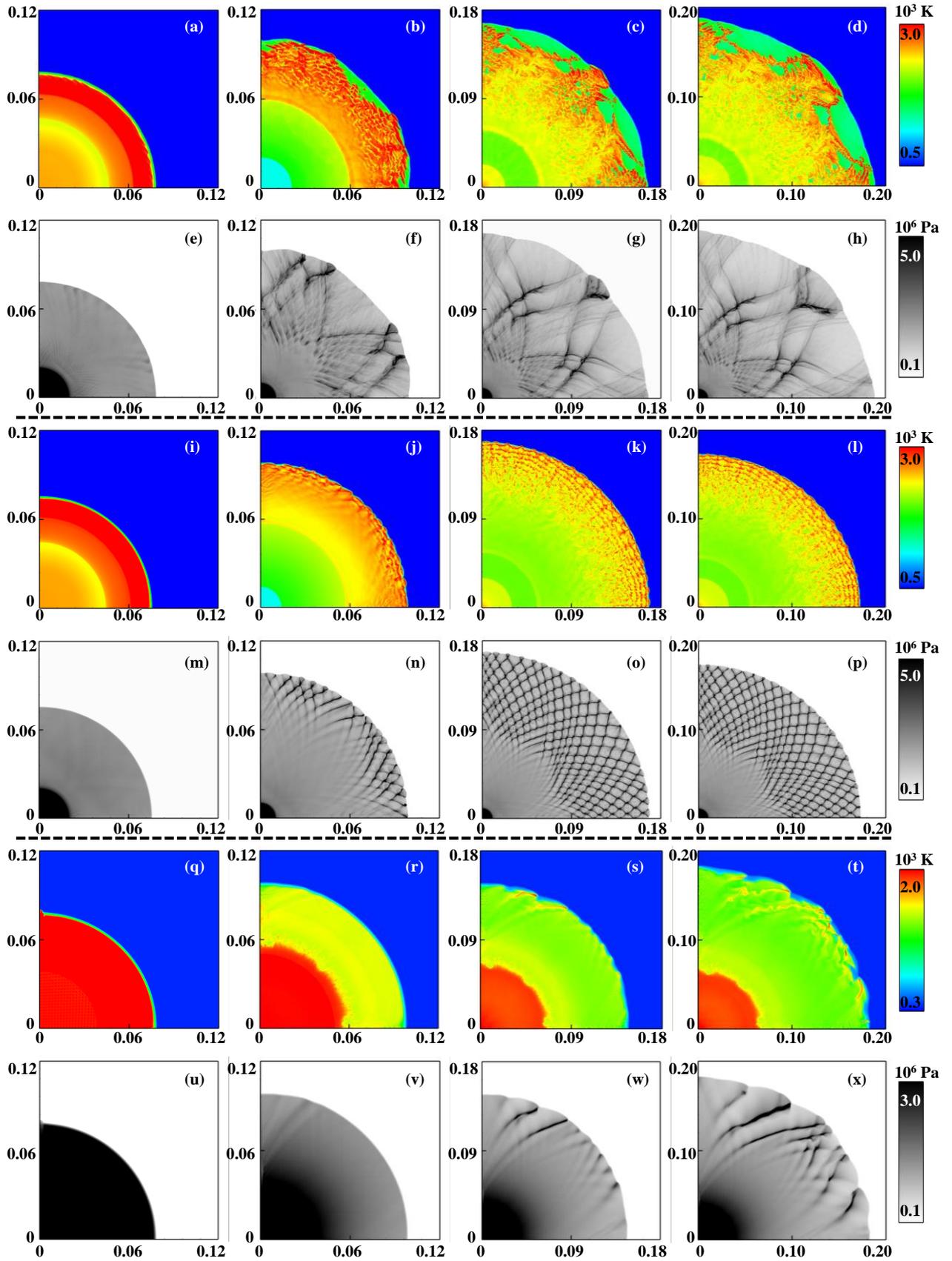

Figure 20: Contours of temperature and peak pressure trajectories: (a)-(h) for $c_0 = 15$ g/m$^3$, (i)-(p) for $c_0 = 200$ g/m$^3$, (q)-(x) for $c_0 = 1{,}750$ g/m$^3$. $d_p = 1$ μm and $\phi = 1$.

In Fig. 18(b), the stable DDI case with $c_0 = 200$ g/m$^3$ is characterised by three conditions:

overdriven, stable, and pulsating. Following the overdriven detonation phase ($R < 0.04$ m), the detonation wave enters a stable development phase within the region $0.04 \leq R \leq 0.07$ m, with nearly constant peak pressure. The comparison of temperature distribution and detonation cells across different particle concentrations in Fig. 20 also shows higher temperatures and regular cell structures at $c_0 = 200$ g/m$^3$, indicative of stable detonation propagation. Subsequently, the detonation fully develops, leading to pressure pulsations.

When $c_0 = 1{,}750$ g/m$^3$, the blast wave decays rapidly (see Fig. 18c), resulting in a cell-free state. Specifically, a pulsating detonation occurs in the region $R > 0.14$ m, with the LSF speed falling below half the C–J speed ($\approx 700$ m/s). This behaviour resembles that of a fast-propagating or choking regime flame in gas mixtures, as noted by Gamezo et al. [61] and Ciccarelli et al. [62]. Notably, Figs. 20(q–t) show that the distance between the LSF and RF remains constant during detonation propagation, indicating minimal decoupling between them. Furthermore, distinct cell structures are absent, with several transverse waves present instead. The temperature and pressure levels in this case are lower than in the other two cases, alongside the weakening effect of particles on the LSF speed. A finer Cartesian mesh (25×25 μm$^2$) was also employed to verify the mesh-independence of the cell-free detonation structure, which can be found from section J of the supplementary document. This is due to the less exothermic and more pronounced endothermic behaviour of high-concentration particles, as detailed in Section 3.1. Generally, as particle concentration increases further, DDI failure occurs, similar to the observations at $c_0 = 2{,}000$ g/m$^3$, as discussed in Fig. 5(f).

The temperature evolutions of homogeneous and heterogeneous reactions are analysed for various particle concentrations. Figure 21(a) shows the spatial temperature profiles of gas and particles along the diagonal direction for $c_0 = 200$ g/m$^3$. The results indicate that the temperatures of the gas and particles in the two-phase mixtures are closely aligned. This is consistent with the temperature contours in Figs. 15(a) and 15(d), which is primarily due to the high efficiency of heat exchange between the gas and particles for small particle sizes. Figure 21(b) presents the circumferential average of gas and particle temperatures after successful detonation initiation at R = 0.175 m, where the detonation waves are fully developed. Particle concentrations ranging from 15 to 1,500 g/m³ are considered. It is observed that the temperatures of both gas and particles are similar, generally increasing initially and then decreasing as particle concentration increases. Both gas and particle temperatures follow an inverted U-shaped trend with varying particle concentrations, which can be attributed to the dual competitive effects of the particles, as discussed in Section 3.1.

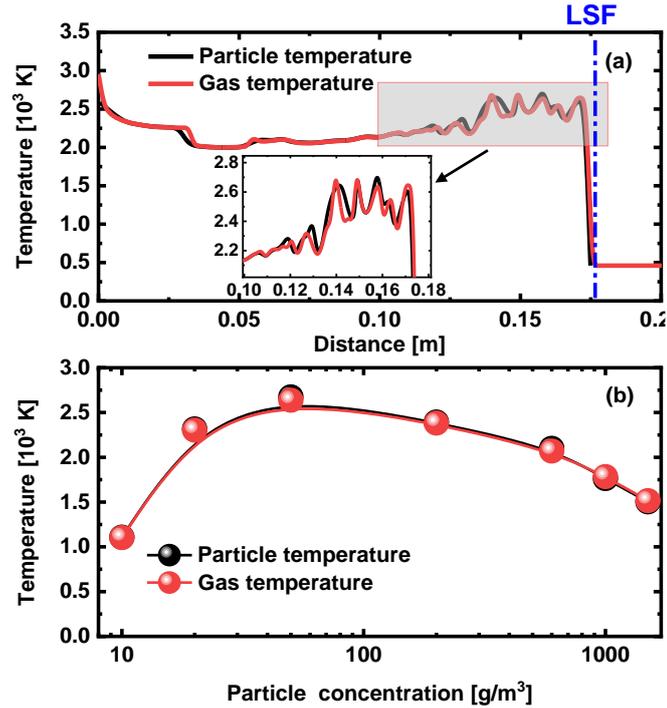

Figure 21: (a) Spatial profiles of the gas temperature and particle temperature for $c_0$ = 200 g/m³, LSF: leading shock front; (b) circumferential average of gas temperature and particle temperature for $R$ =0.175 m and $c_0$ = 15-1,500 g/m³.

**4.3 Interactions between homogeneous and heterogeneous reactions**

Figure 22 shows the heat release rates of homogeneous reactions ($\dot{Q}_{hm}$), heterogeneous reactions ($\dot{Q}_{ht}$), and the interphase energy transfer rate ($S_e$).s The profiles of $\dot{Q}_{hm}$, $\dot{Q}_{ht}$, and $S_e$ along the diagonal direction (Line 1 in Fig. 22a) are presented in Fig. 23. Note that a positive energy exchange rate ($S_e > 0$) indicates energy transfer from particles to gas. Figure 23(a) shows that $S_e$ is approximately zero before the blast wave ($R > 0.012$ m in this case), reflecting rapid energy transfer due to the small coal particle size, which leads to nearly complete energy transfer before the blast wave arrives.

To further explain the energy exchange as the detonation wave propagates outward, we first examine the region 0.009 m < $R$ ≤ 0.012 m (between Line 2 and Line 3), which corresponds to post-detonation flows. The energy exchange in the gas-solid mixtures primarily consists of two conditions behind the detonation front. Firstly, $S_e < 0$ is observed at 0.0114 m < $R$ ≤ 0.0122 m. This occurs because gas-phase reactions take place before particle surface reactions, with a peak heat release of $\dot{Q}_{hm,max} \approx 7.7 \times 10^{12}$ J/m³/s at $R \approx 0.0117$ m (see Fig. 23b), heating the coal particles. Subsequently, surface reactions occur, with a peak heat release of $\dot{Q}_{ht,max} \approx 3.2 \times 10^{12}$ J/m³/s at $R \approx 0.0113$ m (Fig. 23c), and the high-temperature particles continuously release heat to the gas. This corresponds to a relatively long zone with $S_e \geq 0$.

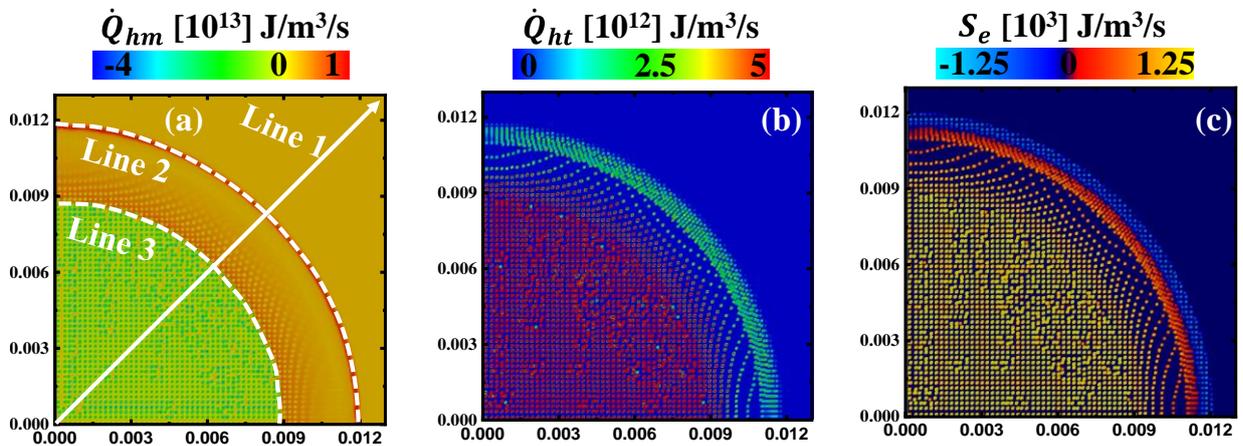

Figure 22. Distributions of (a) heat release rate of homogeneous reaction, (b) heat release rate of heterogeneous reaction, and (c) gas-solid heat exchange. $c_0 = 200$ g/m³, $d_p = 1$ μm, and $\phi = 1$. Unit for the coordinate axis is m.

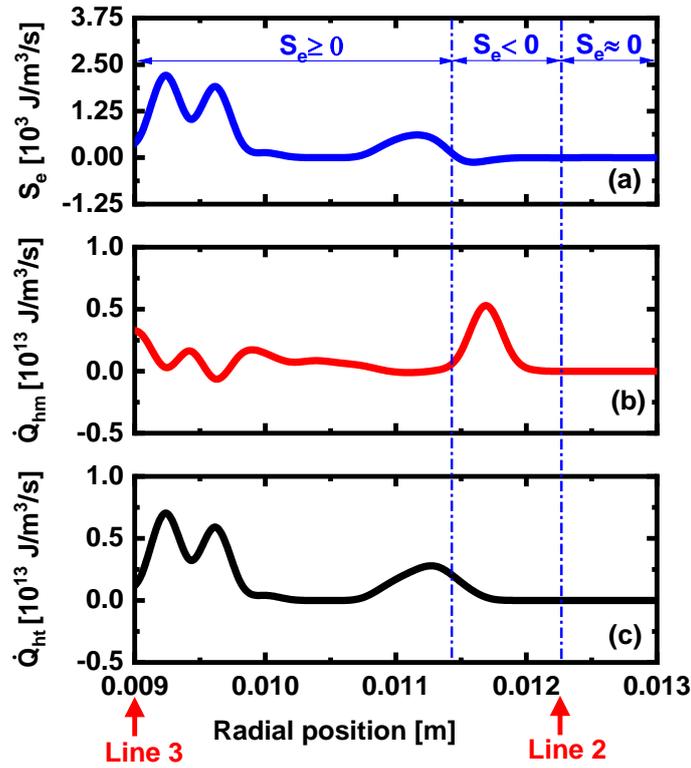

Figure 23: Radial profiles of (a) interphase heat exchange, (b) heat release rate of homogeneous reaction, and (c) heat release rate of heterogeneous reaction.

To analyse the effects of particle addition and surface reactions on methane combustion, we focus on the heat release from gas reactions under specific particle conditions ($c_0$ = 200 g/m$^3$ and $d_p$ = 1 μm) and a unity gas equivalence ratio ($\phi$ = 1). Figure 24(a) shows that the heat release from gas reactions comprises two components. The first is driven by homogeneous reactions, resulting in a heat release rate of approximately 4×10$^{11}$ J/m$^3$/s. During this process, around 20% of the methane is consumed (see Fig. 24c). Additionally, the heat released from particles ($\dot{Q}_{ht}$) exhibits multiple peaks due to the non-uniform combustion of heterogeneous reactions. After this initial gas reaction, the temperature and pressure rise to approximately 2,000 K and 2 MPa, respectively, as shown in Fig. 24(b). The second gas reaction then occurs, characterised by a negative heat release rate, indicating heat absorption. This continues until all oxygen is consumed, at which point the heat release from the particles ceases. Meanwhile, heat absorption from the gas phase gradually decreases, as illustrated in Fig. 24(a).

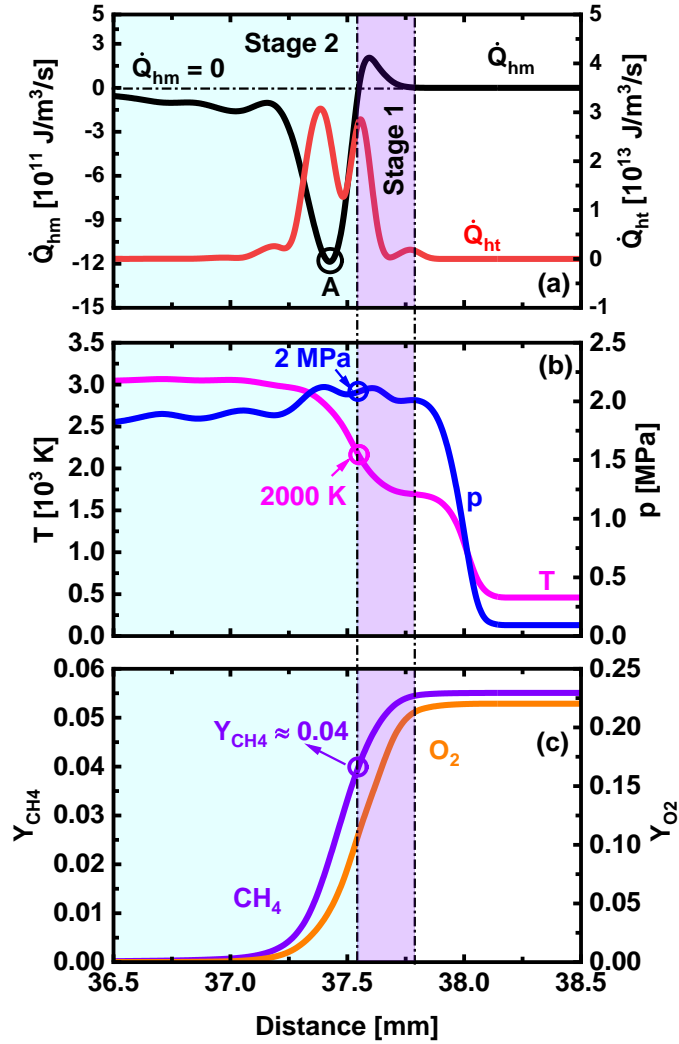

Figure 24: Profiles of distance evolution along the diagonal line of the domain. (a) heat release from homogeneous and heterogeneous reactions, $\dot{Q}_{hm}$, $\dot{Q}_{ht}$; (b) pressure, $p$, and temperature, $T$; (c) mass fractions of methane and oxygen, $Y_{CH_4}$, $Y_{O_2}$. $c_0 = 200$ g/m$^3$, $d_p = 1$ μm, and $\phi = 1$.

## 5. Conclusions

This study numerically investigates the DDI and detonation propagation of two-phase cylindrical detonation in methane/coal particles/air mixtures utilizing the Eulerian-Lagrangian approach. Two-dimensional configuration is considered. A detailed chemical mechanism for methane combustion is employed, whilst a one-step reaction is used for particle surface reaction. A comprehensive analysis has been made on the influences of particle concentration, particle diameter, and gas equivalence ratio on the initiation, propagation, and structure of two-phase detonation. Key conclusions are summarized as below:

(1) A thorough analysis has been performed on the effects of particle concentration, particle

diameter, and gas equivalence ratio. The initiation and propagation of detonation are characterized by LSF speed, peak pressure trajectories, and maximum pressure evolution. The results show that small particle (e.g., $d_p$ = 1 μm) facilitate the DDI. The optimal conditions for DDI occur at a particle concentration of 200 g/m3. The larger particle concentration, particle diameter, and gas equivalence ratio make it more challenging to initiate detonation.

(2) In methane/coal particles/air mixtures, the behaviours of detonation initiation demonstrate intricate dynamics due to the presence of particles. Specifically, the detonation development can be generally categorized into inert ash, burning particles, and unburned particles, based on the dispersed particles. Three DDI modes, i.e., the critical, stable, and cell-free detonations, are identified with increasing particle concentration, revealing a dual impact of particles on DDI. Moreover, the temperatures of the gas and particles are close in the detonated mixtures. Both gas and particle temperatures initially increase and then decrease as the particle concentration increases. Additionally, the purely particle DDI is successful, characterized by greater intensity and instability.

(3) The interactions between homogeneous and heterogeneous reactions are analysed as the detonation wave propagates outwardly. During the energy transfer exchange, gas phase releases heat first, followed by the particles. It also found that the heat release from gas reactions consists of two parts due to the addition of the particles. The first is driven by homogeneous reactions, while another negative heat release rate owing to the exothermic of heterogeneous reactions. This study furthers the theoretical understanding of direct detonation initiation in methane/coal particles/air mixtures by providing a comprehensive analysis.

**Acknowledgement**

This work used the computational resources of the National Supercomputing Centre, Singapore supported by (https://www.nscc.sg/). SL is supported by the China Scholarship Council (202108210264), KG is National Natural Science Foundation of China (52274205), and HZ is supported by Singapore Ministry of Education Grant (A-8000199-00-00).